\documentstyle[12pt]{article}
\begin{document}

\title{\hfill{\normalsize Preprint INRNE-TH-93/4 (May 1993)}\\[-3mm]
       \hfill{\normalsize e-print quant-ph/0001028}\\[5mm]    
            GENERALIZED  INTELLIGENT  STATES AND \\
                $SU(1,1)$  AND $SU(2)$ SQUEEZING\footnote
    {This preprint was sent [with the here preserved mis-spellings Heizenberg, 
studed, ..., and the false degeneracy of the eigenvalue of $L(\lambda)$] 
to Phys. Rev.  Lett.  in May 1993 (LF5064/ 03 Jun 93) and 
declined from PRL in August 1993. An extended version of it appeared later in 
J. Math. Phys. {\bf 35}, 2297 (1994). Meanwhile similar (but not all) 
results were published by other authors in PRL and Phys. Rev. A.}}

\author{                           D.A. Trifonov \\
       Institute for Nuclear Research and Nuclear Energy\\
       Blv. Tzarigradsko chaussee 72, 1784 Sofia, Bulgaria }
\maketitle

\begin{abstract}
    A sufficient condition for a state $|\psi\rangle$ to minimize the
Robertson-Schr\"{o}dinger uncertainty relation for two observables $A$
and $B$ is obtained which for $A$ with no discrete spectrum is also a
necessary one. Such states, called generalized intelligent states (GIS),
exhibit arbitrarily strong squeezing (after Eberly) of $A$ and $B$.
Systems of GIS for the $SU(1,1)$ and $SU(2)$ groups are constructed and
discussed. It is shown that $SU(1,1)$ GIS contain all the Perelomov
coherent states (CS) and the Barut and Girardello CS  while the Bloch
CS are subset of $SU(2)$ GIS.

PACS' numbers 03.65.Ca; 03.65.Fd; 42.50.Dv .
\end{abstract}

\section{Introduction} 

     The squeezed states of electromagnetic field in which the
fluctuations in one of the quadrature components $Q$ and $P$ of
the photon annihilation operator $a = (Q + iP)/ \sqrt{2}$ are
smaller than those in the ground state $|0\rangle$ have atracted
due attention in the last decade (see for example the review
papers\cite{1,2} and references there in). In the recent years an interest
is devoted to the squeezed states for other
observables\cite{3}--\cite{11}. One looks for non gaussian states
which exhibit $Q$-$P$ squeezing\cite{3}--\cite{7} and/or for states in
which the fluctuations of other physical observables are
squeezed\cite{7}--\cite{11}.

   The aim of the present paper is to construct $SU(1,1)$ and $SU(2)$
squeezed intelligent states and to consider some general properties of
squeezing for an arbitrary pair of quantum observables $A$ and $B$ in
states which minimize the Robertson-Schr\"odinger uncertainty relation
(R-S UR)\cite{12}. We call such states generalized intelligent states (GIS)
or squeezed intelligent states when the accent is on their squeezing
properties. The $Q$-$P$ GIS are well studed and known as squeezed states,
two photon coherent states (CS) (see references in\cite{{1},{2}}),
correlated states\cite{13} or Schr\"{o}dinger minimum uncertainty
states\cite{14}. The term intelligent states (IS)\cite{11} is refered to
states that provide the equality in the Heizenberg UR for $A$ and $B$.
The $Q$-$P$ IS are also known as Heizenberg minimum uncertainty states. The
spin IS are introduced and studed in\cite{11}.

\section{Generalized intelligent states} 

     For any two quantum observables $A$  and $B$  the  corresponding
second momenta in a given state obey the R-S UR\cite{{12},{13}},
\begin{equation}\label{UR}                      
  \sigma_{A}^{2} \, \sigma_{B}^{2} \geq \frac{1}{4}
(\langle C \rangle^{2} + 4\sigma_{AB}^{2}),\quad C \equiv -i[A,B] ,
\end{equation}
where $\sigma_{A},  \sigma_{B}$ and $\sigma_{AB}$ are the dispersions and
the covariation of $A$ and $B$,
\begin{eqnarray}\label{2}                       
\sigma_{A}^{2}\,=\,\langle A^{2}\rangle - \langle A\rangle^{2},
\nonumber \\
\sigma_{AB} = \frac{1}{2}(\langle AB + BA \rangle) -
\langle A\rangle\langle B\rangle.
\end{eqnarray}
The states that provide the equality in the R-S UR (\ref{UR}) will be called
here generalized intelligent states (GIS). When  the covariation
$\sigma_{AB} = 0$ then the S-R UR coincides with the Heizenberg one. In
paper\cite{13} it was proved that if a pure state $|\psi\rangle$ with
nonvanishing dispersion of the operator $A$ minimizes the R-S UR then it is an
eigenstate of the  operator  $\lambda A + iB$, where $\lambda$ is a complex
number, related to $\langle C \rangle$ and to $\sigma_{i}(\psi), i=A,B,AB.$
Here we prove that  this is a sufficient condition for any state
$|\psi\rangle$.
\newtheorem{guess}{Proposition}  
\begin{guess}\label{p1}
A state $|\psi\rangle$ minimizes the R-S UR (\ref{UR}) if it is an
eigenstate of the  operator  $L(\lambda)  = \lambda A + iB$,
\begin{equation}\label{3}                       
L(\lambda)|z,\lambda\rangle = z|z,\lambda\rangle,
\end{equation}
where the eigenvalue $z$ is a complex number.
\end{guess}

{\em Proof}. Let first  restrict  the  parameter $\lambda$ in
the eigenvalue eqn. (\ref{3}),  $\rm{Re}\,\lambda \neq 0$.  Then we
express $A$ and $B$ in terms of $L(\lambda)$ and $L^{\dagger}(\lambda)$
and obtain
\begin{eqnarray}\label{4}                       
 \sigma_{A}^{2}(z,\lambda)
=\frac{\langle C\rangle}{2\rm{Re}\,\lambda}\,,\qquad
\sigma_{B}^{2}(z,\lambda) =|\lambda|^{2} \frac{\langle C\rangle}
{2\rm{Re}\,\lambda}\,,\nonumber \\
\sigma_{AB}(z,\lambda)  = -\langle C\rangle \frac{\rm{Im}\,\lambda}
{2\rm{Re}\,\lambda}\,,
\end{eqnarray}
where $\langle C\rangle = \langle\lambda,z|C|z,\lambda\rangle$.  The
obtained second momenta (\ref{4}) obey the equality in R-S UR (\ref{UR}).

Let now the eigenvalue equation (\ref{3}) holds for $\rm{Re}\,\lambda  =
0$. This means that the state $|z,\lambda\rangle$ is an eigenstate  of
the Hermitean operator $r A + B$ where  $r = \rm{Im}\,\lambda$. We consider
now the mean value of the non negative  operator  $F^{\dagger}(r)F(r)$,
where $F(r) = rA + B -(r\langle A\rangle + \langle B\rangle)$ and $r$ is
any real number. Herefrom we get the uncertainty relation
\begin{equation}\label{5}                       
\sigma_{A}^{2} \, \sigma_{B}^{2} \geq \sigma_{AB}^{2}\,,
\end{equation}
the equality holding in the eigenstates of $F(r)$ only.
One can consider the equality in (\ref{5}) as the desired equality in the
Robertson-Schr\"odinger UR  if in these states the mean value of the
operator $C$ vanishes. And this is the case. Indeed, consider in
$|z,ir\rangle$ the mean values of the operators $ A(rA + B)$ and $(rA + B)A$.
We easily get the coinsidence of the two mean  values, wherefrom we
obtain $\langle ir,z|C|z,ir\rangle = 0$ .

Thus all eigenstates $|z,\lambda\rangle$ are GIS. One can prove
that  when  the  operator  $A$  has  no  discrete  spectrum   then
for any $|\psi\rangle$ $\sigma_{A}(\psi) \neq 0$, thereby the  condition
(\ref{3})  is  also  necessary  and  all $A$-$B$ GIS (for any $B$)
are  of  the  form $|z,\lambda\rangle$. Such are for example the cases  of
canonical  $Q$-$P$ GIS\cite{14} and the $SU(1,1)$ GIS, considered
below. The above result stems from the following property of the
dispersion of quantum observables:
\begin{equation}\label{6a}
\sigma_{A}(\psi) = 0  \Longleftrightarrow  A|\psi\rangle =
 a|\psi\rangle.
\end{equation}

   As a consequence of the second part of the proof of the
Proposition~\ref{p1}  we have the following
\begin{guess}\label{p2}
If the commutator $C = -i[A,B]$ is a positive  operator  then  the
operator $rA + B$ with real $r$ has no eigenstates in the  Hilbert
space.
\end{guess}
  In terms of GIS  $|z,\lambda \rangle$ the above Proposition~\ref{p2}
gives the restriction on $\lambda$: $\rm{Re}\,\lambda \neq  0  $  in
cases of positive $C$.

   Before going to  examples let us point out that the $A$-$B$ IS
$|z,\lambda = 1\rangle \equiv |z\rangle $ are noncorrelated and with equal
variances,
\begin{eqnarray}\label{7}                       
L|z\rangle = z|z\rangle ,\qquad L = L(\lambda = 1) = A + iB , \\
 \label{8}\sigma_{A}^{2}(z)  =  \frac{1}{2}\langle z|C|z\rangle  =
\sigma_{B}^{2}(z).
\end{eqnarray}
We shall call such states  equal  variances  IS  or  non  squeezed
IS, addopting the Eberly and Wodkiewicz\cite{7} definition of $A$-$B$
squeezed states. It is convenient to describe this squeezing by means of
the  dimensionless  parameter $q_{A}$\cite{8}
\begin{equation}\label{8a}
q_{A} = \frac{\langle C\rangle/2 \,- \,\sigma_{A}^{2}} {\langle C\rangle/2} ,
\end{equation}
in terms of which the 100\% squeezing corresponds to $q_{A} =  1$.
In the equal variances IS $|z\rangle$  $q_{A} = 0 = q_{B}$.

Let now consider the cases when the
commutator $C=-i[A,B]$ is a positive operator: $\langle\psi|C|\psi\rangle >
0 $. In such cases $\rm{Re}\lambda \neq 0 $ and we can safely devide by
$\langle\psi|C|\psi\rangle$.  Then
from eqns (\ref{4}) we get the quite general result for squeezing
in GIS $|z,\lambda\rangle$ with positive $C$,
\begin{equation}\label{8b}
q_{A}(z,\lambda) = 1 - \frac{1}{2\rm{Re}\,\lambda} ,\qquad
q_{B}(z,\lambda) = 1 - \frac{|\lambda|^{2}}{2\rm{Re}\,\lambda}.
\end{equation}
We see that the squeezing parameter $q$ depends on $\lambda$  only
and  100\%  squeezing  of  $A$  is  obtained  at   $\rm{Re}\,\lambda
\rightarrow \infty $ (and of $B$ at $\lambda = 0$).

     In many cases the IS $|z\rangle$ are  constructed.  Except  of  the
canonical $Q$-$P$ case we point out also the cases of  lowering  and
raising operators of some semisimple Lie  groups  (the $SU(2)$
and the $SU(1,1)$\cite{15} for example) and for the quantum group
$SU(1,1)_{q}$, constructed recently\cite{10}. The GIS $|z,\lambda \rangle$
are eigenstates of the linearly transformed operator
\begin{equation}\label{9}
L  \longrightarrow  L(\lambda) = uL + vL^{\dagger},
\end{equation}
where $u = (\lambda + 1)/2$, $v = (\lambda - 1)/2$,  $L^{\dagger} =
A - iB $. If this is a similarity transformation then GIS can be obtained
by acting on $|z\rangle$ with the transforming operator $S(\lambda)$ (the
generalized squeezing operator) as it was done by Stoler (see the reference
in\cite{1,2}) in the canonical case. In the examples below we construct GIS
by solving the eigenvalue equations of $L(\lambda)$.

\section{$SU(1,1)$ squeezed intelligent states}
     In this section we construct and discuss  $K_{1}$-$K_{2}$  GIS,
where $K_{1}$ and $K_{2}$  are  the  generators  of  the  discrete
series $D^{+}(k)$ of representations  of  $SU(1,1)$  with  Cazimir
operator $C_{2}  :=  k(k-1)$. From the commutation relation $
[K_{1},K_{2}] = -iK_{3}$ we see
that one can apply the corresponding formulas  of  the  previous
section with $A = K_{1}, B = -K_{2}$ and $C = K_{3}$. The operator
$K_{3}$ is positive with eigenvalues $k + m$ where $  m  =  0,1,2,
\ldots,$ . Then as a consequence of the Proposition~\ref{p2} the  GIS
$|z,\lambda;k\rangle$ exist only if $\rm{Re}\,\lambda \neq 0$  and  one  can
safely use formulas (\ref{4}) for the second momenta of $K_{1,2}$
in the $SU(1,1)$ GIS $|z,\lambda;k\rangle$. Since the operator $K_{1}$ has
no discrete spectrum the condition (\ref{3}) is also necessary for
GIS.

     The $SU(1,1)$ equal variances IS $|z;k\rangle$  (the  eigenstates  of
$ K_{1} -iK_{2}$ $\equiv K_{-}$) have been
constructed and studed by Barut and Girardello as `new ``coherent''
states associated with noncompact groups'\cite{15}.
These states form an overcomplete family of states and  provide  a
representation of any state $|\psi\rangle$ in terms of entire  annalytic
function $\langle\psi|z;k\rangle$ of $z$ of order 1  and  type 1  (exponential
type). In the Hilbert space of such entire analytic functions the
generators of $SU(1,1)$ act as the following differential operators
\cite{15} (we shall call this BG-representation)
\begin{eqnarray}\label{12}                     
K_{3} = k + z\frac{d}{dz}\,,\quad K_{+} = K_{-}^{\dagger} = z\,, \nonumber \\
K_{-} = 2k\frac{d}{dz} + z\frac{d^{2}}{dz^{2}}\,.
\end{eqnarray}
   We use the BG-representation to construct the  $SU(1,1)$  GIS
$|z^{\prime},\lambda;k\rangle$ (we denote for a while the eigenvalue by
$z^{\prime}$). The eigenvalue equation (\ref{3}) now reads
\begin{equation}\label{13}                      
\left[u(2k\frac{d}{dz} + z\frac{d^{2}}{dz^{2}}) + vz\right]
\Phi_{z^{\prime}}(z) = z^{\prime}\Phi_{z^{\prime}}(z) \,,
\end{equation}
where the parameters $u,v$ have been defined in formula (\ref{9}).
By means of a simple substitutions the above equation is reduced
to the Kummer equation for the confluent  hypergeometric  function
$_{1}F_{1}(a,b;z)$ \cite{16}, so that we have  the  following  solution
of eqn. (\ref{13})
\begin{eqnarray}\label{14}                      
\Phi_{z^{\prime}}(z) = \exp{(cz)}\, _{1}F_{1}(a,b;-2cz)\,, \\
a = k-\frac{z^{\prime}}{2uc}\,,\quad b = 2k;\quad c^{2} = -\frac{v}{u}\,.
\end{eqnarray}
This solution obey the requirements of the BG representation iff
\begin{equation}\label{16}                           
|c| = \sqrt{|v/u|} < 1 \Leftrightarrow \rm{Re}\,\lambda
> 0\,,
\end{equation}
which is exactly the  restriction on $\lambda$ imposed by the
positivity of the commutator $C \equiv K_{3}$, according to the
Proposition~\ref{p2}. No other constrains on $z^{\prime}$ and
$\lambda$  are needed. Thus we obtain the $SU(1,1)$ GIS $|z^{\prime},
\lambda;k\rangle$ in the BG-representation in the form
\begin{equation}\label{17}                            
\langle k;\lambda,z^{\prime}|z;k\rangle = \exp{(c^{*}z)}\, _{1}F_{1}
(a^{*},b;-2c^{*}z)\, ,
\end{equation}
where  the  parameters  $a,b$  and  $c$  are  given  by   formulas
(3.4). Using  the  power series of $_{1}F_{1} (a,b;z) $\cite{16} we
get the coinsidence of our solution (\ref{17}) at $\lambda = 1$
($u = 1$, $v=0$) with the solution of Barut and Girardello\cite{15},
\begin{equation}\label{18}                            
\langle k;\lambda=1,z^{\prime}|z;k\rangle =
{}_{0}F_{1}(2k;z{z^{\prime}}^{*})
\,=\, \langle k;z^{\prime}|z;k\rangle .
\end{equation}

We note the twofold degeneracy of the eigenvalues of the operator
$L(\lambda \neq 1)$ as it is seen from  eqn. (3.4). We denote the
two solutions as $\langle\pm;k;\lambda,z^{\prime}|z;k\rangle$.
The degeneracy is removed at $\lambda = 1$ as it is known from the
BG-solution. Thus this point is a branching point for the operator
$L(\lambda)$. It worth noting that the degeneracy is also removed by the
following constrain on the two complex parameters $z^{\prime}$ and $\lambda$
in eqn. (3.6)
\begin{equation}\label{19}                            
z^{\prime} \, = \, 2k\sqrt{-uv} \,= \,k\sqrt{1-\lambda^{2}}\,.
\end{equation}
Using the properties of the function $_{1}F_{1}(a,b;z)$ \cite{16} we
get from (\ref{17}) in both $(\pm)$ cases the same expression
$\exp{(z\sqrt{-v^{*}/u^{*}})}$ which can be seen to be nothing but the
BG-representation of the Perelomov $SU(1,1)$ CS $|\zeta;k\rangle$\cite{17}
with $\zeta = \sqrt{-v/u}$\,\,,
\begin{equation}\label{20}                            
 |\zeta;k\rangle = (1-|\zeta|^{2})^{k}\,\exp{(\zeta K_{+})}
 \,|k;k\rangle \,.
\end{equation}
If we impose $z^{\prime} = -2k\sqrt{-uv}$ we get CS$|-\zeta;k\rangle$.
One can directly check (using the $SU(1,1)$ commutation relations
only) that CS (\ref{20}) are indeed eigenstates of $L(\lambda)$, eqn.
(\ref{9}), with eigenvalue (\ref{19}) provided $\zeta^{2} = -v/u$ .
We calculate  explicitly the first and second momenta of the generators
$K_{i}$ in CS $|\zeta;k\rangle$ (for $\sigma_{K_{i}}$ see also\cite{8})
\begin{eqnarray}\label{21,22}                        
\sigma_{K_{1}K_{2}}= -2k\,\frac{\rm{Re}\,\zeta\,\rm{Im}\,\zeta}
{(1-|\zeta|^{2})^{2}}\,, \nonumber \\
\sigma_{K_{1}}^{2} = \frac{k}{2}\,\frac{|1+\zeta^{2}|^{2}}
{(1-|\zeta|^{2})^{2}},\quad
\sigma_{K_{2}}^{2} = \frac{k}{2}\,\frac{|1-\zeta^{2}|^{2}}
{(1-|\zeta|^{2})^{2}}
\end{eqnarray}
and convince that the equality in the R-S UR (\ref{UR}) is satisfied.

Thus all the Perelomov $SU(1,1)$ CS are GIS. They are represented
by the points of the two dimensional surface (\ref{19}) in the
four dimensional space of points $(z,\lambda)$. The BG CS\cite{15} form
another subset of $SU(1,1)$ GIS isomorfic to the plane $\lambda = 1$.

   We note that the aboved formulas for the first and second momenta of
$K_{i}$ in CS $|\zeta;k\rangle$ hold also for the (non square integrable)
Lipkin-Cohen representation with Bargman index $k = 1/4$ (but not for
$k= 3/4$ ),
\begin{eqnarray}\label{23}                           
K_{1}  \,=\, \frac{1}{4}\,(Q^{2}-P^{2}),  \quad K_{2}
\,=\, -\frac{1}{4}(QP+PQ), \nonumber \\
K_{3} \,=\, \frac{1}{4}(Q^{2}+P^{2}) .
\end{eqnarray}
Due to the expressions of $K_{i}$ in terms of the canonical pair $Q,P$
the CS $|\zeta;k=1/2,1/4,3/4\rangle$ ($|\zeta;k=1/4,3/4\rangle$ are
eigenstates of the squared boson operator $a^{2}$)
are of interest for $Q$-$P$ squeezing\cite{4,14,18}. One can also
calculate the fluctuations of $Q$ and $P$\cite{18} and show
that CS $|\zeta;k=1/4\rangle$ exhibit about 56\% ordinary
squeezing (Bu\v{z}ek\cite{4}). The squeezing of $K_{1,2}$ in CS
$|\zeta;k\rangle$ has been studed in\cite{8}: the 100\% squeezing (in the
sense of the parameter $q$, eqn. (\ref{8a}) for $K_{1}$ is obtained at
$\zeta  =  i$. We note however that
$$\sigma_{i}^{2}(\zeta;k)\, \geq \, \frac{k}{2} \,= \,\sigma_{i}^{2}
(0;k) , \quad i =  K_{1},K_{2} \, ,$$
i.e. no squeezing of $\sigma_{i} $ in $|\zeta;k\rangle$ in comparison with
the ground state $|0;k\rangle$.

  In conclusion to this section we note that for $SU(1,1)$  GIS
the squeezing operator $S(\lambda)$ exists and can be defined by means of
the relation $|z,\lambda;k\rangle = S(\lambda)|z;k\rangle$ since the spectra
of $L$ and $L(\lambda)$ coinside. It belongs again to
the $SU(1,1)$ (but not to the series $D^{+}(k)$ since one can show
that it is not unitary) and its matrix elements $\langle k;z|S|z;k\rangle$
are explicitly given by the functions (\ref{17}) with $z^{\prime} = z$. These
diagonal matrix elements determine $S$ uniquely due to the  analyticity
property of the BG-representation\cite{15}. We recall that the same property
of the diagonal matrix elements holds in the canonical (Glauber) CS
representation (see for example\cite{2} and references therein).

\section{$SU(2)$ squeezed intelligent states}
     Let now $A,B$ and $C$ be the generators $J_{1}, -J_{2}$ and
$-J_{3}$ of $SU(2)$ group, i.e. the spin operators of spin
$j  = 1/2,1, \ldots,$. In this example the commutator $C=-J_{3}$ is not
positive (the limit $\rm{Re}\,\lambda = 0$ can be taken) and the
operator $A = J_{1}$ has a disctete spectrum (some of its eigenstates
are examples of exceptional GIS which are not eigenstates of
$L(\lambda)$). In paper\cite{11} there were constructed the
eigenstates (in their notations) $|w_{N}(\tau)\rangle$ of the operator
$J(\alpha)=J_{1}-i\alpha J_{2},$ where $N=0,1,2 \ldots, 2j$, $\tau^{2} =
(1-\alpha)/(1+\alpha)$, $\alpha $ being arbirary complex number. These
states are eigenstates also of $L(\lambda)  =  \lambda J_{1} - iJ_{2}  $,
thereby they all are $J_{1}$-$J_{2}$ GIS, minimizing the R-S UR
(\ref{UR}).  They can be represented in the general form
$|z_{N},\lambda;j\rangle$ with the eigenvalues $z_{N}$ $ =
(j-N)\sqrt{\lambda^{2}-1}$. Among them (for $N=0$ and $N=2j$) are
the Bloch (the spin or the $SU(2)$) CS $|\tau;-j\rangle$ and
$|-\tau;-j\rangle$ ($\tau $ is any complex number)
\begin{equation}\label{24}                          
|\tau;-j\rangle = (1+|\tau|^{2})^{-j}\exp{(\tau J_{+})}|-j\rangle.
\end{equation}

The mean values of $J_{i}, i=1,2,3$ and $J_{i}^{2}$ (and the
dispersions $\sigma_{J_{1}}$ and $\sigma_{J_{2}}$) in Bloch CS are
known\cite{11,19}. Calculating also the covariation,
\begin{equation}\label{25}                           
\sigma_{J_{1},J_{2}}(\tau)  =  2j\frac{\rm{Re}\,\tau \,\rm{Im}\,\tau}
{(1+|\tau|^{2})^{2}}
\end{equation}
we can directly check that in CS $|\tau\rangle$ the equality in the R-S
UR (\ref{UR}) holds for the spin operators  $J_{1,2}$. Thus the
Bloch CS are a subset of the $SU(2)$ GIS.

     Let us briefly discuss the properties of the $SU(2)$ GIS.
First of all for a given parameter $\lambda$ there are $2j+1$
independent GIS $|z_{N},\lambda;j\rangle$.  There is only one equal
variances IS, namely $|-j\rangle$, the point $\lambda = 1$ being again
the branching point of the $L(\lambda)$. From this fact it follows
that squeezing operator does not exist. Since the commutator $C=
i[J_{1},J_{2}] = -J_{3}$ the limit $\rm{Re}\,\lambda  =  0$ in GIS
is alowed and in the fluctuations formulas (\ref{4}) as well since
at this limit $\langle C\rangle=\langle J_{3}\rangle = 0$. The operator
$A = J_{1}$ has a discrete spectrum, therefore $\sigma_{A} \geq 0$. From the
explicit formula
\begin{equation}\label{26}                            
\sigma_{J_{1}}^{2}(\tau)  =  \frac{j}{2}\frac{|1-\tau^{2}|^{2}}
{(1+|\tau|^{2})^{2}}
\end{equation}
we see that this fluctuation vanishes at $\tau^{2} = 1$. Therefore
in virture of the property (\ref{6a}) the Bloch CS $|\tau=\pm1;-j\rangle$
are eigenstates of $J_{1}$ which can be checked also directly, the
eigenvalues being $\pm j$. The other eigenstates of $J_{1}$ are exactly
those exceptional states which minimize the R-S UR (\ref{UR}) but are not
of the form $|z,\lambda\rangle$ (i.e. dont obey eqn.(\ref{3})). The final
note we make about  $SU(2)$ GIS is that except for the eigenvalue $z_{N}=0$
(when $N=j$) all the others are not degenerate (unlike the $SU(1,1)$ case).
\section{Concluding remarks}
We have presented a method for construction of squeezed intelligent
states (called here generalized intelligent states (GIS))
for any two quantum observables $A$ and $B$ in which 100\%
squeezing (after Eberly) can be obtained. GIS minimize the
Robertson-Schr\"{o}dinger uncertainty relation and can be considered
as a generalization of the canonical $Q$-$P$ squeezd states\cite{13}.
When the operators $A$ and/or $B$ are exspressed in terms of the canonical
pair $Q,P$ one can look in the $A$-$B$ GIS for the squeezing of $Q$ end/or
$P$ as well. Such are for example the cases of $SU(1,1)$ GIS for the
representations with Bargman indexes $k=1/4, 1/2,3/4$. The $SU(1,1)$
GIS form a larger set of states which contains as two different subsets the
Perelomov CS and the Barut and Girrardello CS.

   The method is based on the minimization of the Robertson-Schr\"{o}dinger
UR (\ref{UR}) for which the eigenvalue equation (\ref{3}) for the operator
$L(\lambda)=\lambda A +iB$ is a sufficient condition. In case of $A$ with
continuous spectrum this is also a necessary conditon independently on $B$.
In view of this the method provides the possibility (when one is interested
in squeezing of the fluctuations of $A$) to look for the best squeezing
partner of $A$. Thus for example if $A=P$ then one can show that the
eigenstates of $L(\lambda)$ exist for a series $B = Q^{n}$,
$n = 1,5,9,\ldots,$.

   When the $A$-$B$ GIS can be obtained from the equal variances IS
$|z\rangle$  by means of the invertable squeezing operator $S(\lambda)$
the latter belongs to $SU(1,1)$ as it can be derived from (\ref{9}).
This fact shows that
$SU(1,1)$ plays important role in a wide class of squeezing phenomina
(not only in $Q$-$P$ case).

\subsection*{Acknowledgments}
This work is partialy supported by Bulgarian Science
Foundation research grant \# F-116.


\begin{thebibliography}{99}

\bibitem{1} R.~Loudon and P.~Knight, J. Mod. Opt. {\bf 34}, 709 (1987).
\bibitem{2} W.~Zhang, D.~Feng and R.~Gilmore. Rev. Mod. Phys. {\bf 62},
867 (1990).
\bibitem{3} G.~D'Ariano, M.~Rasetti and M.~Vadacchino, Phys. Rev. D
{\bf 32}, 1034 (1985).
\bibitem{4} V.~Bu\v{z}ek. J. Mod. Opt. {\bf 37}, 159 (1990); J.~Sun et
all. Phys. Rev. A{\bf 44}, 3369 (1991); C.~Gerry and E.~Hach III, Phys.
Lett. A{\bf 174}, 185 (1993).
\bibitem{5} J.~Katriel et all, Phys. Rev. D{\bf 34}, 2332 (1986).
\bibitem{6} P.~Kral, J. Mod. Opt. {\bf 37}, 889 (1990).
\bibitem{7} K.~Wodkiewicz and J.~Eberly, J. Opt. Soc. Am. B{\bf 2},
458 (1985); K.~Wodkiewicz, J. Mod. Opt. {\bf 34}, 941 (1987).
\bibitem{8} V.~Bu\v{z}ek, J. Mod. Opt. {\bf 37}, 303 (1990).
\bibitem{9} J.~Vaccaro and D.~Pegg, J. Mod. Opt. {\bf 37}, 17 (1990).
\bibitem{10} L.~Kuang and F.~Wang, Phys. Lett. A{\bf 173}, 221 (1993).
\bibitem{11} C.~Aragone, E~Chalband and S.~Salamo, J. Math. Phys. {\bf
17}, 1963 (1976).
\bibitem{12}  H.~Robertson, Phys. Rev. {\bf 35}, 667 (1930);
S.~Schr\"{o}dinger, Ber. Kil. Acad. Wiss., s. 296, Berlin (1930).
\bibitem{13} V.~Dodonov, E.~Kurmyshev and V.~Man'ko, Phys. Lett. A{\bf
76}, 150 (1980).
\bibitem{14} D.~A.~Trifonov, J. Math. Phys. {\bf 34}, 100 (1993).
\bibitem{15} A.~O.~Barut and L.~Girardello, Commun. Math.  Phys.  {\bf
21}, 41 (1971).
\bibitem{16} {\it Handbook on Mathematical Functions}, edited by
M.~Abramowitz and I.~A.~Stegun (National Bureau of Standarts, 1964;
Russian translation, Nauka, 1979).
\bibitem{17} A.~M~Perelomov, Commun. Math. Phys. {\bf 26}, 222 (1972).
\bibitem{18} B.~A.~Nikolov and D.~A.~Trifonov, Commun. JINR. E2-81-798
(Dubna, 1981).
\bibitem{19} E.~H.~Lieb, Commun. Math. Phys. {\bf 31}, 327 (1973).
%
\end{thebibliography}
\end{document}